\documentclass[12pt,aps,prb,preprint]{revtex4}
\usepackage{amsmath}    
\usepackage{graphicx}   
\usepackage{bm}

\begin{document}

\draft 
\title{Relativistic addition of perpendicular velocity components from the constancy of the speed of light}

\author{Ben Yu-Kuang Hu}
\altaffiliation[On sabbatical at ]{Department of Physics,
University of Maryland, College Park, MD 20742-4111}
\altaffiliation[Electronic mail: ] {benhu@physics.uakron.edu}
\affiliation{Department of Physics, University of Akron, Akron,
OH~44325-4001}

\date{\today}

\begin{abstract}
Mermin [Am. J. Phys. {\bf 51}, 1130--1131 (1983)] derived the
relativistic addition of the parallel components of velocity using
the constancy of the speed of light. In this note, the derivation
is extended to the perpendicular components of velocity.
\end{abstract}
\maketitle

Mermin\cite{mermin83} gave a succinct elementary derivation of the
relativistic addition the parallel component of the velocity using
the constancy of the speed of light, without resorting to Lorentz
transformations.  Here, I show that his derivation can be extended
to the addition of the perpendicular components of the velocity.

As in Ref.~\onlinecite{mermin83}, we consider a train moving on a
long straight track at constant speed $v$.  At a certain instant,
a photon, which has speed $c$, and a massive particle, which has
speed less than $c$, are both projected from one side the train,
say the right side, in the direction perpendicular (according to
someone standing on the train) to the direction of motion of the
train. When the photon reaches the left side of the train, it is
immediately reflected back.  On its way back to the right side, it
encounters the massive particle at some point.

Since the place on the train where the photon and the massive
particle meet is frame-independent, the ratio $r$ of the
perpendicular components of the velocity of the massive particle,
$w_\perp$, and that of the photon must be an invariant. To a
person standing on the ground, the parallel component of the
photon's velocity is $v$, the speed of the train. Since the speed
of the photon is constant at $c$ in any frame, the perpendicular
component of the velocity according to the person standing on the
ground is, by Pythagoras' theorem, $\sqrt{c^2 - v^2}$. Therefore,
the invariant ratio is
\begin{equation}
r = \frac{w_\perp}{\sqrt{c^2 - v^2}}\;.
\end{equation}

Consider two cases for the speed of the train, $v$ and $v'$.  The
invariance of $r$ implies
\begin{eqnarray}
\frac{w_{\perp}}{\sqrt{c^2 - v^2}} &=&
\frac{w_{\perp}'}{\sqrt{c^2 - v'^2}}\nonumber\\
\Rightarrow   w_{\perp}' &=& w_{\perp}\sqrt{\frac{1- v'^2/c^2}{1 -
v^2/c^2}}. \label{eq:const}
\end{eqnarray}
All that is left to do is to express $v'$ in terms of the velocity
difference $V$ between $v'$ and $v$ using the parallel velocity
addition rule\cite{mermin83}
\begin{equation}
v' = \frac{v + V}{1 + v V/c^2}.
\end{equation}
Substituting this into Eq.~(\ref{eq:const}) gives the desired
result
\begin{equation}
w_{\perp}' = w_{\perp} \frac{\sqrt{1-V^2/c^2}}{1 + v V/c^2}.
\end{equation}

\end{document}